\documentclass[a4paper,fleqn,usenatbib]{mnras}

\usepackage[T1]{fontenc}
\usepackage{ae,aecompl}

\usepackage{graphicx}	% Including figure files
\usepackage{amsmath}	% Advanced maths commands
\usepackage{amssymb}	% Extra maths symbols
\usepackage{multirow}

\newcommand{\chandra}{\textit{Chandra}}
\newcommand{\nustar}{\textit{NuSTAR}}
\newcommand{\suzaku}{{\it Suzaku}}
\newcommand{\swift}{{\it Swift}}
\newcommand{\xmm}{{\it XMM-Newton}}

\title[Mrk~766 with \nustar]{\nustar\ observations of Mrk~766: distinguishing reflection from absorption}

\author[D. J. K. Buisson et al.]{D. J. K. Buisson$^{1}$,\thanks{Email: djkb2@ast.cam.ac.uk}
	M. L. Parker$^{1,2}$,
	E. Kara$^{3}$,
	R. V. Vasudevan$^{1}$,
	A. M. Lohfink${^4}$,
\newauthor	C. Pinto${^1}$,
	A. C. Fabian$^{1}$,
   D. R. Ballantyne$^{5}$,
   S. E. Boggs$^{6}$,
   F. E. Christensen$^{7}$,
\newauthor   W. W. Craig$^{7,8}$,
   D. Farrah$^{9,10}$,
   C. J. Hailey$^{11}$,
   F. A. Harrison$^{12}$,
   C. Ricci$^{13,14,15}$,
\newauthor   D. Stern$^{16}$,
   D. J. Walton$^{1}$ and
   W. W. Zhang$^{17}$\\
  $^{1}$Institute of Astronomy, Madingley Road, Cambridge, CB3 0HA\\
  $^{2}$European Space Astronomy Centre (ESA/ESAC), E-28691 Villanueva de la Ca\~nada, Madrid, Spain\\
  $^{3}$Department of Astronomy, University of Maryland, College Park, MD 20742-2421, USA\\
  $^{4}$Montana State University, Bozeman, 59717-3840, MT, USA\\
  $^{5}$Center for Relativistic Astrophysics, School of Physics, Georgia Institute of Technology, 837 State Street, Atlanta, GA 30332-0430, USA\\
  $^{6}$Space Sciences Laboratory, University of California, Berkeley, CA 94720-7450, USA\\
  $^{7}$DTU Space-National Space Institute, Technical University of Denmark, Elektrovej 327, DK-2800 Lyngby, Denmark\\
  $^{8}$Lawrence Livermore National Laboratory, Livermore, CA 94550, USA\\
  $^{9}$Department of Physics and Astronomy, University of Hawaii, 2505 Correa Road, Honolulu, HI 96822, USA\\
  $^{10}$Institute for Astronomy, 2680 Woodlawn Drive, University of Hawaii, Honolulu, HI 96822, USA\\
  $^{11}$Columbia Astrophysics Laboratory, Columbia University, New York, NY 10027, USA\\
  $^{12}$Cahill Center for Astrophysics, 1216 E. California Blvd, California Institute of Technology, Pasadena, CA 91125, USA\\
  $^{13}$N\'ucleo de Astronom\'ia de la Facultad de Ingenier\'ia, Universidad Diego Portales, Av. Ej\'ercito Libertador 441, Santiago, Chile\\
  $^{14}$Kavli Institute for Astronomy and Astrophysics, Peking University, Beijing 100871, China\\
  $^{15}$Chinese Academy of Sciences South America Center for Astronomy, Camino El Observatorio 1515, Las Condes, Santiago, Chile\\
  $^{16}$Jet Propulsion Laboratory, California Institute of Technology, 4800 Oak Grove Drive, Mail Stop 169-221, Pasadena, CA 91109, USA\\
  $^{17}$X-ray Astrophysics Laboratory, NASA Goddard Space Flight Center, Greenbelt, MD 20771, USA
}

\date{Accepted 30$^{\rm th}$ July 2018. Received 27$^{\rm th}$ July 2018; in original form 23$^{\rm rd}$ May 2017}

\pubyear{2018}

\begin{document}
\label{firstpage}
\pagerange{\pageref{firstpage}--\pageref{lastpage}}
\maketitle

\begin{abstract}
We present two new \nustar\ observations of the narrow line Seyfert 1 (NLS1) galaxy Mrk~766 and give constraints on the two scenarios previously proposed to explain its spectrum and that of other NLS1s: relativistic reflection and partial covering.
The \nustar\ spectra show a strong hard ($>15$\,keV) X-ray excess, while
simultaneous soft X-ray coverage of one of the observations provided by \xmm\ constrains the ionised absorption in the source.
The pure reflection model requires a black hole of high spin ($a>0.92$) viewed at a moderate inclination ($i=46_{-4}^{+1}\,^\circ$).  
The pure partial covering model requires extreme parameters:
the cut-off of the primary continuum is very low ($22_{-5}^{+7}$\,keV) in one observation and the intrinsic X-ray emission must provide a large fraction (75\%) of the bolometric luminosity. 
Allowing a hybrid model with both partial covering and reflection provides more reasonable absorption parameters and relaxes the constraints on reflection parameters.
The fractional variability reduces around the iron~K band and at high energies including the Compton hump, suggesting that the reflected emission is less variable than the continuum.
\end{abstract}

\section{Introduction}

A common feature of the X-ray spectra of many non-jetted active galactic nuclei (AGN) is the hard excess, a strong increase in flux above $\sim15$~keV. This was first detected in stacked spectra from {\it Ginga} \citep{Pounds90} and measured for individual sources with {\it BeppoSAX} \citep{perola02}. Prior to the launch of \nustar , detailed measurements had only been made in a handful of AGN, using \swift-BAT or the \suzaku\ PIN detector, which were interpreted either as evidence of Compton thick absorption \citep{Risaliti09,Turner09} or the Compton hump of reflected emission \citep{Walton10}. X-ray reflection \citep{lightman88,George91} occurs when the primary X-ray source, known as the corona, illuminates the accretion disc or other relatively cold material such as the torus. This illumination triggers the emission of fluorescent lines at low energies and is scattered into a `Compton hump' at high energies. When the reflection spectrum originates from the parts of the accretion disc close to the innermost stable circular orbit (ISCO) of the black hole, the narrow features are blurred out by relativistic effects \citep{Fabian89, Laor91}.  

Since the launch of \nustar\ \citep{Harrison13}, the Compton hump has been more definitively detected in many AGN \citep{Risaliti13,Walton14,Marinucci14,Brenneman14,Parker14c,Balokovic15,Kara15}. The sensitivity of \nustar\ at high energies means that it can be used to differentiate between reflection and absorption models for the hard excess \citep[e.g.][]{vasudevan14}.

One object which has had both these processes proposed to explain its spectrum is Mrk~766. Mrk~766 is a nearby ($z=0.013$) narrow line Seyfert 1 (NLS1) galaxy. NLS1s are thought to be rapidly accreting ($\dot{m}\sim0.01-1$), relatively low mass (typically $M_{\rm BH}\sim10^{6}-10^{7}M_{\odot}$) AGN, and are distinguished by narrow optical Balmer lines, weak [O \textsc{iii}] and strong Fe \textsc{ii} emission \citep[see review by][]{Komossa08}. In the X-ray band, NLS1s are spectrally soft, and are thus easily detected by low energy instruments. They frequently show complex, rapid variability and non-trivial spectral shapes, so are of great interest for study. 
The supermassive black hole in the nucleus of Mrk~766 has a mass of 1--6$\times10^6 M_\odot$ \citep{Bentz09,Bentz10} and the host is a barred spiral galaxy.
Spectrally, the evidence for a relativistically-broadened iron line in Mrk~766 is tentative. A broad line was claimed with {\it ASCA} by \citet{Nandra97asca}. However, later analysis of a more sensitive \xmm\ spectrum by \citet{Pounds03} showed that the line profile could instead be described by ionized reflection alone, with no need for relativistic blurring. Based on \xmm\ and \suzaku\ observations of Mrk~766, \citet{Miller07} and \citet{Turner07} proposed a model where the bulk of the spectral variability is due to variations in multiple complex (partially-covering, ionized) absorbing zones. A recent re-analysis of the archival \xmm\ data by \citet{Liebmann14} showed that the spectra and variability could be well described by a composite model, containing both partial-covering absorption and relativistic reflection. 

More robust evidence for the presence of relativistic reflection in Mrk~766 comes from the detection of a reverberation lag \citep{Emman11,DeMarco13}, thought to be caused by the time delay induced in the reflected signal due to the light travel time from the corona to the disc. \citet{Emman11} found almost identical reverberation lags in Mrk~766 and MCG--6-30-15, the first source in which a broad iron line was discovered \citep{Tanaka95}. Mrk~766 is included in the sample of objects studied by \citet{Emman14}, who found that by modelling the time lag spectra they could precisely determine the mass ($M_{\rm BH}=1.6_{-1.2}^{+1.4}\times 10^{6}\,M_{\sun}$) and constrain other physical parameters (e.g. the dimensionless spin, $a>0.56$). The discovery of iron K lags in some sources \citep[e.g.][]{zoghbi12,Kara13,kara16}, which have so far only been explained by invoking relativistic reflection, have reinforced the interpretation of these high-frequency time lags as originating from reverberation close to the black hole. However, iron K reverberation lags have not yet been detected in Mrk~766 \citep{kara16}.

In this paper, we present the results of recent \nustar\ observations of Mrk~766, where we examine the hard X-ray spectrum using the sensitivity and high-energy spectral resolution of \nustar\ to enable us to constrain the different physical models for the hard excess.
The observations and methods of data reduction are presented in Section~\ref{section_datareduction}; results of the analysis are given in Section~\ref{sec:results}; these results are discussed in Section~\ref{section_discussion}; and conclusions are made in Section~\ref{section_conclusions}.

\section{Observations and Data Reduction}
\label{section_datareduction}

\begin{table}
\caption{List of \nustar\ observations of Mrk~766 and associated simultaneous X-ray observations.}
\label{tab:obs}
\begin{tabular}{lrcr}
\hline
Telescope & OBSID & Start time & \multirow{1}{1.5cm}{Observation length/ks}\\
\\
\hline
\nustar &60101022002& 2015-01-24T12:31 & 90.2 \\
\swift-XRT  &00080076002& 2015-01-25T00:08 &  4.9 \\
\hline
\nustar &60001048002& 2015-07-05T22:24 & 23.6 \\
{\it XMM}-EPIC  &0763790401 & 2015-07-05T17:26 & 28.2 \\

\hline

\end{tabular}

\end{table}

Mrk~766 has been observed twice by \textit{NuSTAR}: for 90\,ks starting on 2015 Jan 24 and for 23\,ks starting on 2015 July 5. The first observation had a simultaneous \textit{Swift} snapshot and the second was taken jointly with \textit{XMM-Newton} (see Table~\ref{tab:obs}).

The \nustar\ data were reduced using the \nustar\ data analysis software (NuSTARDAS) version 1.4.1, and CALDB version 20140414. We extracted cleaned event files using the \textsc{nupipeline} command, and spectral products using the \textsc{nuproducts} command, using 80\,arcsec radius circular extraction regions for both source and background spectra. The background region was selected from a region on the same chip, uncontaminated with source photons or background sources.

The \swift\ data were reduced using the \swift\ XRT products generator, using the procedure described in \citet{Evans09} to extract a spectrum.

The \xmm\ data, taken in small window mode, were reduced according to the standard guidelines in the
\xmm\ User's Manual, using the \xmm\ \textsc{Science
Analysis Software (sas)} version 14.0.0 as described in \citet{vasudevan13}. The task \textsc{epchain} 
was used to reduce the data from the pn
instrument. An annular source region of
outer radius 40\,arcsec and inner radius 10\,arcsec was used to extract a source spectrum to remove mild pileup (detected using the \textsc{epatplot} tool), checking for
nearby sources in the extraction region. Circular regions near the source were used to calculate the background.
Additionally, the background light
curves (between 10 and 12\,keV) were inspected for flaring,
and a comparison of source and background light curves in
the same energy ranges was used to determine the portions of
the observation in which the background was sufficiently low
compared to the source (4\,ks was lost to flaring); the subsequent spectra were generated
from the usable portions of the observation.
Response matrices and auxiliary files were generated using the
tools \textsc{rmfgen} and \textsc{arfgen}.

We use \xmm-RGS data from the new observation and the highest and lowest flux archival observations (OBSIDs: 0304030101, 0304030301). We reduced the data with the standard \xmm\ pipeline, \textsc{rgsproc}.
We use a source region including 95\% of the PSF and background from outside 98\% of the PSF (\textsc{xpsfincl}=95 and \textsc{xpsfexcl}=98).
The two detectors were added for illustrative purposes only using \textsc{rgscombine}.

Spectra from all instruments were grouped to a signal to noise level of 5. Fits were made in ISIS Version 1.6.2-32 \citep{houck00}; errors are given at the 90\% level.
We use the elemental abundances of \citet{Wilms00} with cross sections from \citet{verner96}.

\section{Results}
\label{sec:results}

\begin{figure}
\centering
\includegraphics[width=\linewidth]{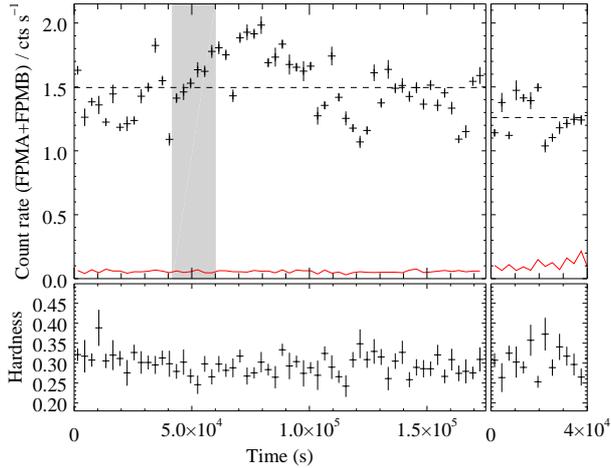}
\caption{Upper Panels: \nustar\ light curve with 3\,ks bins. The gap between panels corresponds to $\sim6$\,months. The background light curve is shown in red, and the time of the \swift\ XRT exposure by the shaded region. The dashed horizontal lines show the mean flux of each observation. Lower panels: 10--40/3--10\,keV hardness ratio.
}
\label{fig_nustarlightcurve}
\end{figure}

From the \nustar\ lightcurve (Fig.~\ref{fig_nustarlightcurve}), we determine that no significant long term flux or hardness variability is seen. It is therefore appropriate to fit average spectra of each observation. The \swift\ X-ray telescope (XRT) snapshot taken during the first \nustar\ observation (shown by the shaded region in Fig.~\ref{fig_nustarlightcurve}) occurred at a flux level close to the average. Therefore, it is likely to be indicative of the average low energy spectrum over the whole observation.

\begin{figure}
\includegraphics[width=\columnwidth]{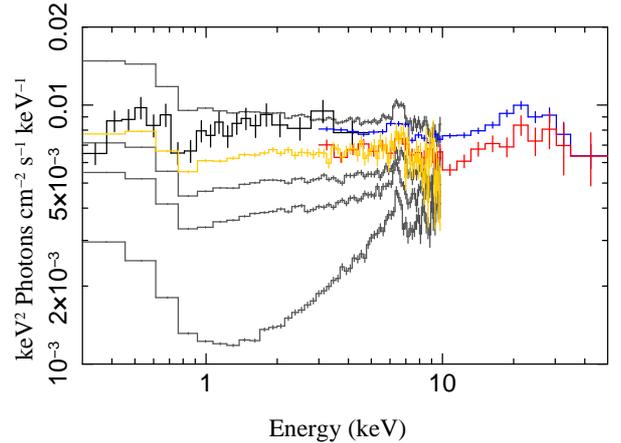}
\caption{Unfolded fluxes (to a $\Gamma=2$ powerlaw) of the January (\textit{NuSTAR}: blue, \textit{Swift}: black) and July (\textit{NuSTAR}: red, \textit{XMM}: orange) observations, with a range of previous \textit{XMM} (grey) observations (OBSIDs 0109141301, 0304030101, 0304030301 and 0304030401, see \citealt{Miller07,xmmmass} for detailed analysis). Spectra have been rebinned for plotting.}
\label{fig:pastcomp}
\end{figure}

We compare the \textit{NuSTAR} observations with previous observations in Fig.~\ref{fig:pastcomp}. This shows that the new observations are close to the high flux end of the previously observed states of Mrk~766.

\begin{figure}
\includegraphics[width=\columnwidth]{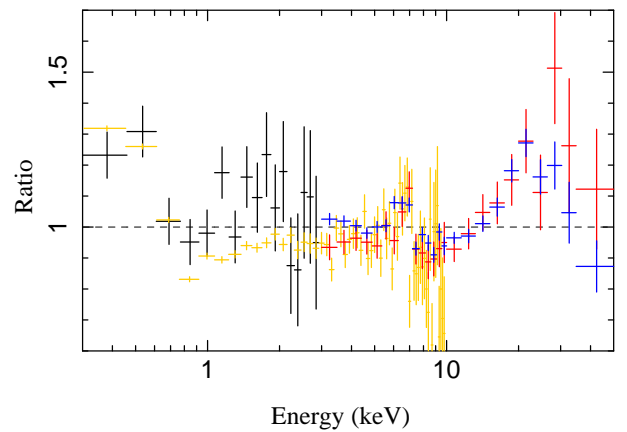}
\caption{Ratios of each observation to a $\Gamma=2$ powerlaw with Galactic absorption: January Swift (black) and NuSTAR (blue); July XMM-PN (orange) and NuSTAR (red). Spectra have been rebinned for plotting.}
\label{fig:nupowres}
\end{figure}

We begin our analysis by comparing all data for each of the new observations to a powerlaw with Galactic absorption (modelled with \texttt{tbnew}, \citealt{Wilms00}).
The residuals to a $\Gamma=2$ powerlaw are shown in Fig.~\ref{fig:nupowres}.
This shows the spectrum is moderately soft, with a soft excess below 0.7\,keV with a deficit above, and excesses at $5-7$\,keV and $15-40$\,keV.

The drop in flux at $\sim0.7$\,keV may be due to warm absorption features such as the OVII edge and the iron unresolved transition array (UTA).
The excesses at $5-7$\,keV and $15-40$\,keV are typical of emission from Fe~K and Compton scattering due to reflection, but similar features can be produced by partial covering absorption reducing the flux at other energies.
We therefore proceed by fitting detailed models of these processes to the spectra.
Since different processes dominate the spectral features in the low and high energy data, we first consider each region separately, before combining these to give a consistent broadband picture.

\subsection{High-energy fits -- iron line and hard excess}
\label{sub:nufit}

To constrain the iron line and hard excess while minimising the effect of absorption, we initially fit the data above 3\,keV. Since the \swift\ observation has little signal above 3\,keV, we fit only the data from \nustar\ and \textit{XMM}-pn. We fit the two observations separately.

\begin{figure*}
\includegraphics[width=\linewidth]{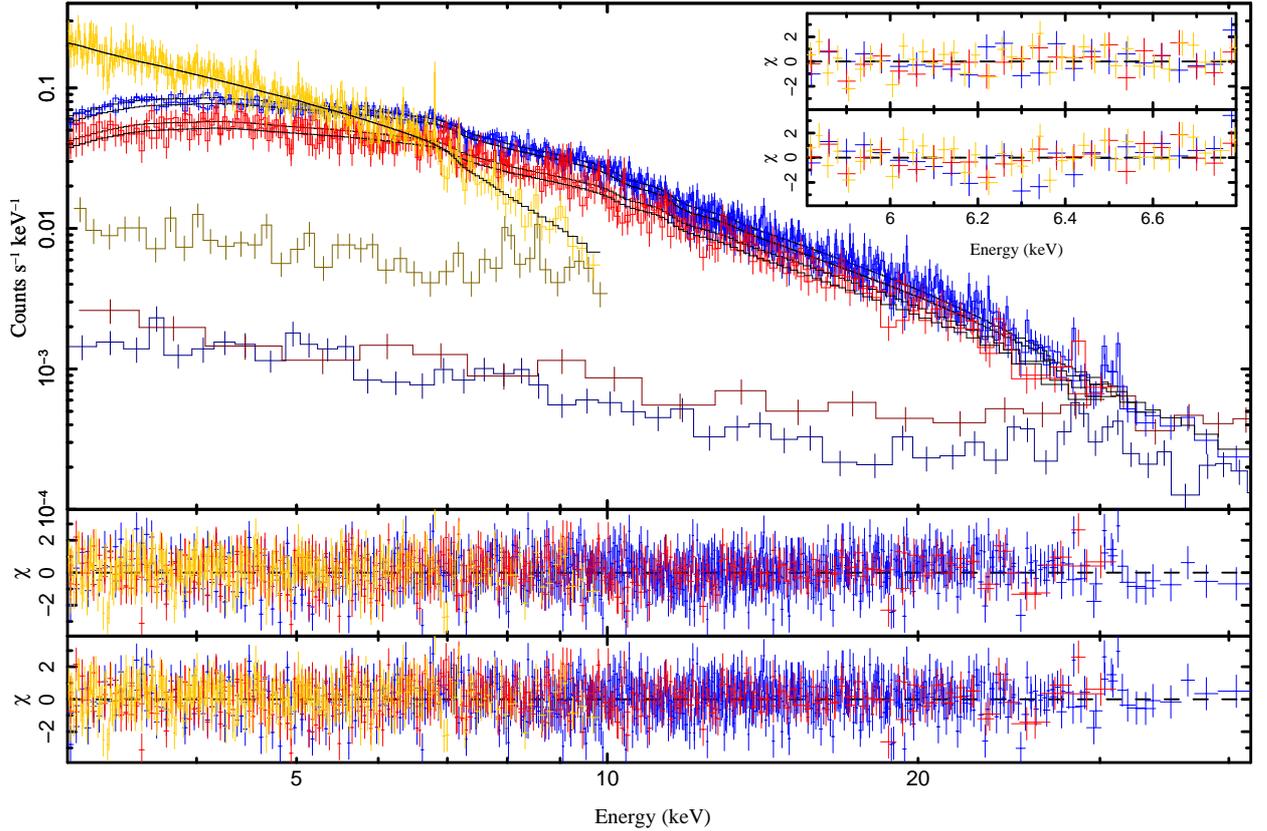}
\caption{Top panel: fits to \nustar\ and \xmm\ data above 3\,keV (the model shown is of relativistic reflection). \xmm\ data in orange is simultaneous with the red (dimmer) \nustar\ data; the other \nustar\ observation is shown in blue. Backgrounds are shown in dark orange, red and blue respectively. Middle panel: residuals from relativistic reflection model. Lower panel: residuals from partial covering model.
Inset: Residuals for reflection models over $5.8-6.8$\,keV. Blurred reflection (upper inset) has smaller residuals around the iron K band than distant neutral reflection (lower inset): over $5.8-6.8$\,keV, $\Delta\chi^2=18$.}
\label{fig:highe_res}
\end{figure*}

In the reflection case, we find that distant reflection (modelled with \texttt{pexmon}, \citealt{magdziarz95,Nandra07}) is insufficient to model the iron line ($\chi^2/{\rm d.o.f.}=1682/1526=1.102$), leaving residuals around the narrow iron line (Fig.~\ref{fig:highe_res}, inset), which suggest that the iron line is broadened.
We test this by replacing the narrow iron line in \texttt{pexmon} with a broadened gaussian (\texttt{pexrav+zgauss}). This gives a significantly better fit, $\Delta\chi^2=46$ and $17$ for the \swift/\nustar\ and \xmm/\nustar\ observations repectively. Parameters of fits to this model are shown in Table~\ref{tab:highe}. This shows significant width to the iron line, likely due to the orbital motion of material around the black hole.

Having shown that the iron line is broadened, we fit the spectrum with a reflection model that incorporates self-consistent relativistic blurring of the entire spectrum (\texttt{relxill}, \citealt{dauser10,garcia14}). This provides a reasonably good fit ($\chi^2/{\rm d.o.f.}=1607/1518=1.059$).
Further, we note that including distant reflection does not provide a significant improvement over relativistic reflection alone ($\Delta\chi^2=6$ for 2 additional degrees of freedom) and no physical parameters of the relativistic model change significantly. Hence, we present models including only relativistic reflection (Table~\ref{tab:highe}).

These models show Mrk~766 as a source with slight relativistic blurring viewed at moderate inclination ($i=42\pm3^\circ$ or ${39_{-3}^{+6}}^{\circ}$). Parameters such as the inclination, which are not expected to change within 6\,months, are consistent between the two observations. The cut-off of the primary continuum is too high to measure.

It has also been suggested that the spectrum of Mrk~766 can be explained by partial covering absorption of the primary source \citep{Miller07,Turner07}. 
We model this with a cut-off power-law with a number of partially covering components, using the model \texttt{zpcfabs}. We find that a model with two components ($\chi^2/{\rm d.o.f.}=1616/1522=1.062$) provides a similar quality fit to the reflection model.
Having only one partial covering component gives a significantly worse fit to the data ($N_{\rm H}\sim9\times10^{24}$\,cm$^{-2}$, $f_{\rm cov}\sim0.6$, $\chi^2/{\rm d.o.f.}=1693/1526=1.110$) and three partial covering components gives insignificant improvement over 2 components ($\Delta\chi^2=8$ for 4 degrees of freedom).

The two-component fit requires a component of strong absorption ($N_{\rm H}>5\times10^{24}$\,cm$^{-2}$) in each observation and a low energy of the cut-off ($E_{\rm cut}=27_{-9}^{+20}$\,keV)  in the \swift/\nustar\ observation, which has better high energy statistics due to the longer \nustar\ exposure.

\begin{table}
\caption{Fits to data from each observation above 3\,keV. Parameters indicated with * are fixed. $R_{\rm Refl}$ indicates the reflection strength, where $R_{\rm Refl}=1$ gives the reflection from material covering $2\pi$\,steradians with an isotropic source.}
\label{tab:highe}
\begin{tabular}{lcc}
\hline
Parameter & \swift/\nustar & \xmm/\nustar \\
\hline
\multicolumn{3}{c}{Reflection with broad Gaussian iron line}\smallskip \\
$\Gamma$ & $2.46_{-0.08}^{+0.08}$ & $2.31\pm0.12$ \\
$E_{\rm cut}$/keV & $>440$ & $>310$ \\
$R_{\rm Refl}$ & $3.5_{-0.9}^{+1.0}$ & $2.1_{-0.9}^{+1.2}$ \\
Line $E$\,/keV & $6.4$*  & $6.4$*\\
Line $\sigma$\,/keV & $2.3_{-0.4}^{+0.3}$  & $1.4_{-0.7}^{+0.5}$\\
$\chi^2/$d.o.f. & $909.2/806=1.128$ & $710.6/718=0.990$ \\
\hline

\multicolumn{3}{c}{Relativistic reflection}\smallskip \\

Emissivity Index & $<4$ & $>4.3$ \\
$a$ & Unconstrained & $<0.44$ \\
$i/^\circ$ & $42_{-3}^{+3}$ & $39_{-3}^{+6}$ \\
$\Gamma$ & $2.24_{-0.05}^{+0.09}$ & $2.16_{-0.08}^{+0.12}$ \\
$\log(\xi/{\rm erg\,cm\,s^{-1}})$ & $1.8_{-0.5}^{+0.4}$ & $<3.1$ \\
$A_{\rm Fe}$ & $1.2_{-0.3}^{+0.7}$ & $2.9_{-1.4}^{+0.8}$ \\
$E_{\rm cut}$/keV & $>210$ & $>230$ \\
$R_{\rm Refl}$ & $1.15_{-0.20}^{+0.25}$ & $1.3_{-0.4}^{+0.5}$\smallskip\\
$\chi^2/$d.o.f. & $902.9/803=1.124$ & $704.1/715=0.985$ \\

\hline
\multicolumn{3}{c}{Partial covering absorption}\smallskip \\

%\hline
$N_{\rm H}/$cm$^{-2}$ & $1.15\pm0.2\times 10^{25}$ & $7.1_{-1.9}^{+2.6}\times10^{24}$\\
$f_{\rm Cov}$ & $0.67_{-0.07}^{+0.06}$  & $0.55_{-0.11}^{+0.15}$\\
$N_{\rm H}/$cm$^{-2}$ & $88_{-15}^{+18}\times10^{23}$ & $4_{-2}^{+3}\times10^{23}$ \\
$f_{\rm Cov}$ & $0.45_{-0.07}^{+0.06}$ & $0.37_{-0.16}^{+0.10}$ \\
$\Gamma$ & $2.2\pm0.15$ & $2.3_{-0.4}^{+0.2}$ \\
$E_{\rm Cut}$/keV & $27_{-9}^{+20}$ & $110_{-90}^{+390}$ \smallskip\\
$\chi^2/$d.o.f. & $909.7/805=1.142$ & $706.0/717=0.996$ \\

\hline

\end{tabular}

\end{table}

\subsection{Low-energy fits - warm absorption and emission}
\label{sub:wabs}

\begin{figure}
\centering
\includegraphics[width=7.8cm]{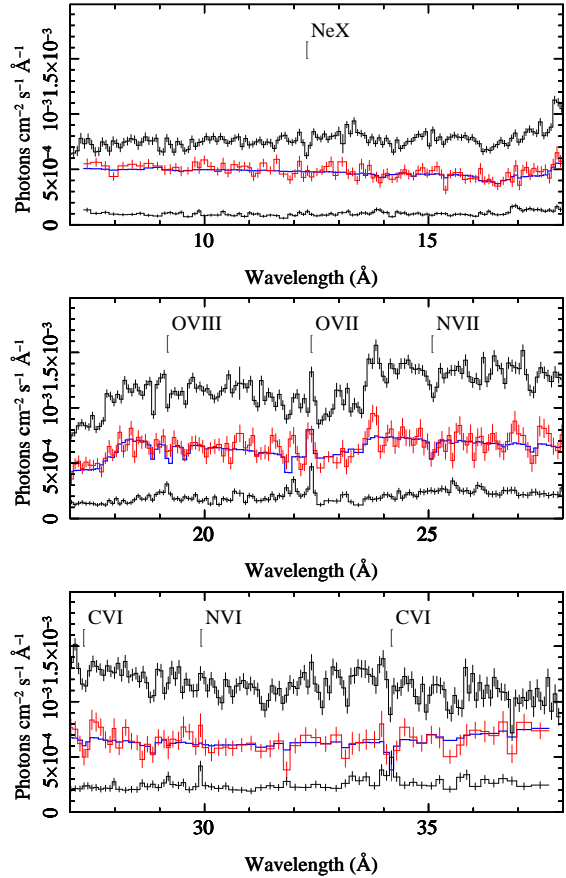}
\caption{
RGS data from the new observation (red) and the highest and lowest archival flux states (black) unfolded to a constant model.
Emission lines 
are clearly visible in the low state, while absorption features are present in the high state.
The blue line shows a fit with absorption and emission components applied to a phenomenological continuum.
Wavelengths are given in the observed frame and the range of each panel overlaps the next by 1\,\AA.}
\label{fig:rgsres}
\end{figure}

\begin{figure*}
\centering
\includegraphics[angle=-90, width=\linewidth]{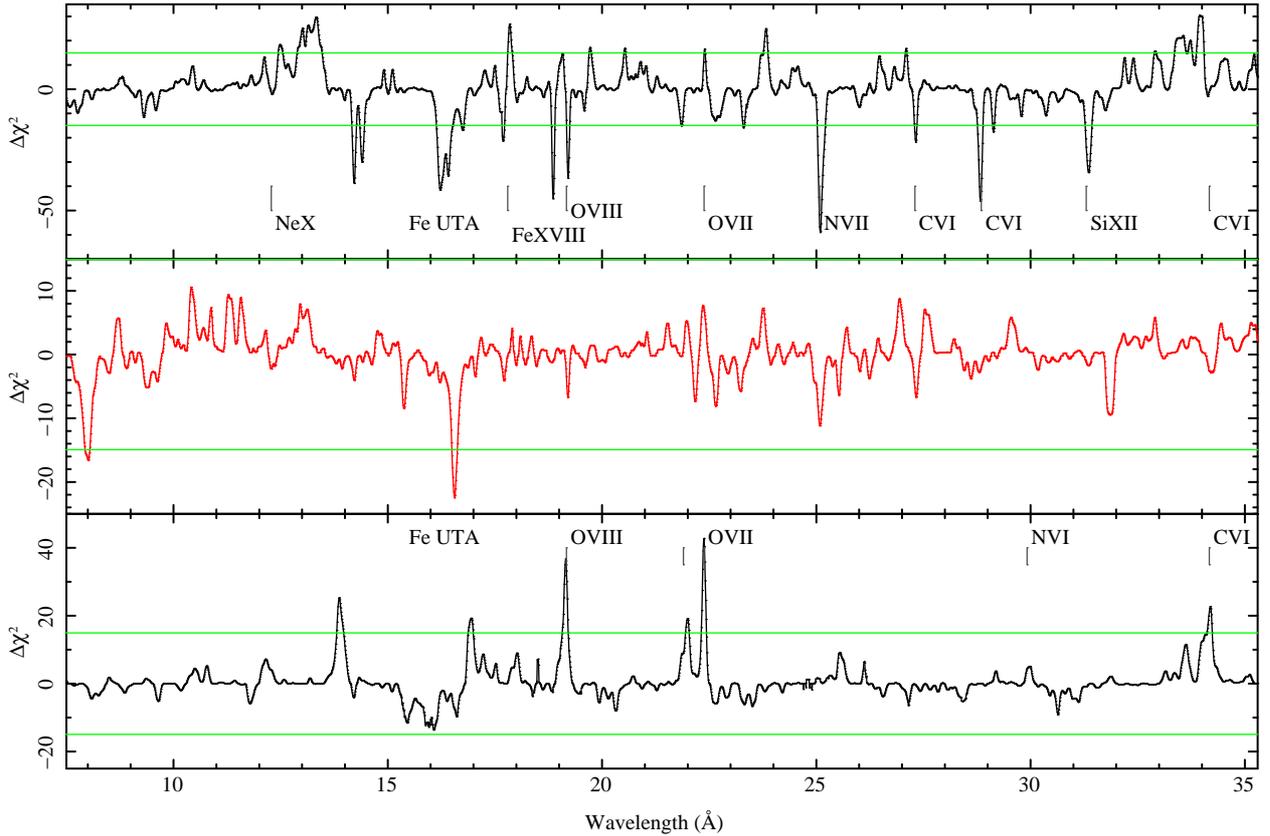}
\caption{Results of line scans to the new data (middle) as well as the highest (top) and lowest (bottom) flux archival observations. The green lines show the 95\,\% confidence interval for a blind search.
}
\label{fig:lines}
\end{figure*}

\begin{table}
\caption{Narrow features in archival RGS spectra. Line wavelengths 
and transition levels are values from the APEC database. Redshifts are given in the observer's frame (Mrk~766 is at $z=0.0129$).}
\label{tab:rgs}

\centering{Low state -- emission}

\begin{tabular}{lrcl}
\hline
\multicolumn{2}{c}{Species} & Rest wavelength/\AA & Redshift \\
\hline
\ion{C}{VI}  & 2p$^1$ -- 1s$^1$ & 33.736  & $0.0130_{-0.0021}^{+0.0008}$ \\ 
\ion{N}{VI}  & 1s$^1$2s$^1$ -- 1s$^2$ & 29.534  & $0.0121_{-0.0004}^{+0.0009}$ \\
\ion{O}{VII} & 1s$^1$2s$^1$ -- 1s$^2$   & 22.098 & $0.0128_{-0.0006}^{+0.0006}$ \\
\ion{O}{VIII}& 2p$^1$ -- 1s$^1$ & 18.969 & $0.0099_{-0.0010}^{+0.0012}$ \\
\hline
\end{tabular}

\bigskip

\centering{High state -- absorption}

\begin{tabular}{lrcl}
\hline
\multicolumn{2}{c}{Species} & Rest wavelength/\AA & Redshift \\
\hline
\ion{Ne}{X} & 1s$^1$ -- 2p$^1$ & 12.134 & $0.0126_{-0.0014}^{+0.0017}$ \\
\ion{N}{VII}& 1s$^1$ -- 2p$^1$ & 24.781 & $0.0132_{-0.0031}^{+0.0009}$ \\
\ion{C}{VI} & 1s$^1$ -- 4p$^1$ & 26.990 & $0.0124_{-0.0006}^{+0.0004}$ \\
\ion{C}{VI} & 1s$^1$ -- 2p$^1$ & 33.736 & $0.0117_{-0.0004}^{+0.0005}$ \\
\hline
\end{tabular}

\end{table}

Warm absorption and emission are known to have an important effect on the spectrum of Mrk~766 in the soft band \citep[e.g.][]{sako03,laha14}. To determine the nature of the gas which is responsible for these features, we begin by identifying features visually and with systematic line scans similar to those performed by for example \citet{tombesi10} and \citet{pinto16}
with a phenomenological continuum (Fig.~\ref{fig:lines}).

We use a broadband continuum model based on that of \citet{rgsrellines}: a powerlaw with two broad lines. We then ensure that the local continuum is well-described with a cubic spline modification over the region $\pm1$\,\AA\ from the wavelength of interest. We measure line significance from the change in $\chi^2$ when including an additional unresolved Gaussian at fixed wavelength, allowed to have positive or negative normalisation (one additional degree of freedom; the use of positive or negative normalisation in the same fit is needed to avoid the problems described in \citealt{protassov02}).
We then scan the wavelength across the RGS range to find $\Delta\chi^2$ at each wavelength. We indicate approximate significance by estimating a critical $\Delta\chi^2_{\rm crit}$ for 95\,\% significance from the expected distribution of $N$ independent trials each having a $\chi_1^2$ distribution. We estimate the number, $N$ of independent trials performed by the wavelength range tested divided by the instrumental resolution, $(\lambda_{\rm max}-\lambda_{\rm min})/{\rm d}\lambda=450$. The global 95\,\% confidence interval (the solution $\Delta\chi^2_{\rm crit}$ to $P(\chi_1^2<\Delta\chi^2_{\rm crit})^N=0.95$) then corresponds to  $\Delta\chi^2_{\rm crit}=14.9$ (shown as the green line in Fig.~\ref{fig:lines}).
We note also that changing the estimate of the effective number of independent trials has only a small effect on the critical $\Delta\chi^2_{\rm crit}$: increasing or decreasing the number of trials by a factor of 2 changes the critical $\Delta\chi^2_{\rm crit}$ to 16.2 or 13.6 respectively.

Since the new observations show few features at high significance (the \xmm~RGS observation is relatively shallow -- 37,000\,counts in total across both detectors and orders), we are also guided by the sensitive archival \xmm~RGS spectra from the highest and lowest previously observed states (Fig.~\ref{fig:rgsres}; pn spectra shown in Fig.~\ref{fig:pastcomp}).
While the absorbing material may not be identical in the \xmm\ and \nustar\ exposures, the features found in the archival observations provide a guide from which to start modelling the latest data.
Results of these line scans are shown in Fig.~\ref{fig:lines}.
Where a feature has a single most likely associated transition, we fit with a Gaussian to find the redshift of the feature. Results are shown in table~\ref{tab:rgs}. The lines are consistent with being unresolved.

The previous low-flux observation shows several emission lines (Table~\ref{tab:rgs}). 
The ionisation states of the observed lines suggest an ionisation parameter of $\log(\xi/{\rm erg\,cm\,s^{-1}})\sim1.5$. Warm emission in AGN is usually predominantly photoionised \citep[e.g.][]{guainazzi07} and the high strength of the \ion{O}{VII} forbidden line relative to the corresponding recombination line in this case supports this. 
Most of the lines are consistent with being in the rest frame of Mrk~766, but the most highly ionised \ion{O}{VIII} line is bluer, with a redshift corresponding to a projected outflow of $900_{-350}^{+300}$\,km\,s$^{-1}$.
We also note that the \ion{N}{VI} line is larger than would be predicted based on Solar abundances. This is not unexpected as overly strong \ion{N}{VI} lines have also been found in NGC~3516 \citep{turner03}.

In contrast, the high-flux spectrum principally shows absorption features (Table~\ref{tab:rgs}), with \ion{O}{VII} the only previously identified emission line still seen in emission (the other emission lines are likely not visible due to their small equivalent width).
 Narrow features are present at wavelengths expected of \ion{C}{VI}, \ion{N}{VII}, and \ion{Ne}{X}.

Since detailed modelling of the warm absorber is not the primary focus of this work, we do not attempt to fit the archival observations but fit the new observations with photoionisation models including the detected features. We use the photoionisation model \texttt{XSTAR} \citep{kallman01} and, for computational efficiency, fit using tables, which we compute for an appropriate region of parameter space.

Fitting the new observation with several warm-absorber components, we find that two ionisation states are sufficient ($\chi^2/{\rm d.o.f.}=1040/982=1.06$).
With only one rather than two absorbing components, the fit is significantly worse ($\Delta\chi^2=36$ for 2 fewer degrees of freedom), as the absorption around the iron unresolved transition array (UTA) region ($\sim17$\,\AA) is not sufficiently broad. Three absorbers provide insignificant improvement ($\Delta\chi^2=1.1$ for 2 additional degrees of freedom).
It is likely that this two-component absorber represents a more complicated region of gas, but this parametrisation is sufficient to describe the absorber well enough to allow broadband continuum fitting.

When modelling the full dataset, we freeze the redshifts to appropriate values due to the large amount of low resolution data, which can drive the fitted redshift away from the values derived from the narrow RGS features. We fix the redshift of the warm absorbers to a value consistent with all the features observed in the high-flux spectrum, $z=0.0118$.
Since the \ion{O}{VIII} line is not detected in the new observation, we model the line emission with a single component of photoionised gas at a redshift consistent with the \ion{O}{VII} line.

\subsection{Broadband fits}

With a description of both the high-energy excesses and the warm absorber separately, we now perform a broadband fit to find a consistent model of the high and low energy features of the spectra. We include all data from \nustar, \swift, \xmm-pn and \xmm-RGS.

\begin{figure*}
\centering
\includegraphics[width=\linewidth]{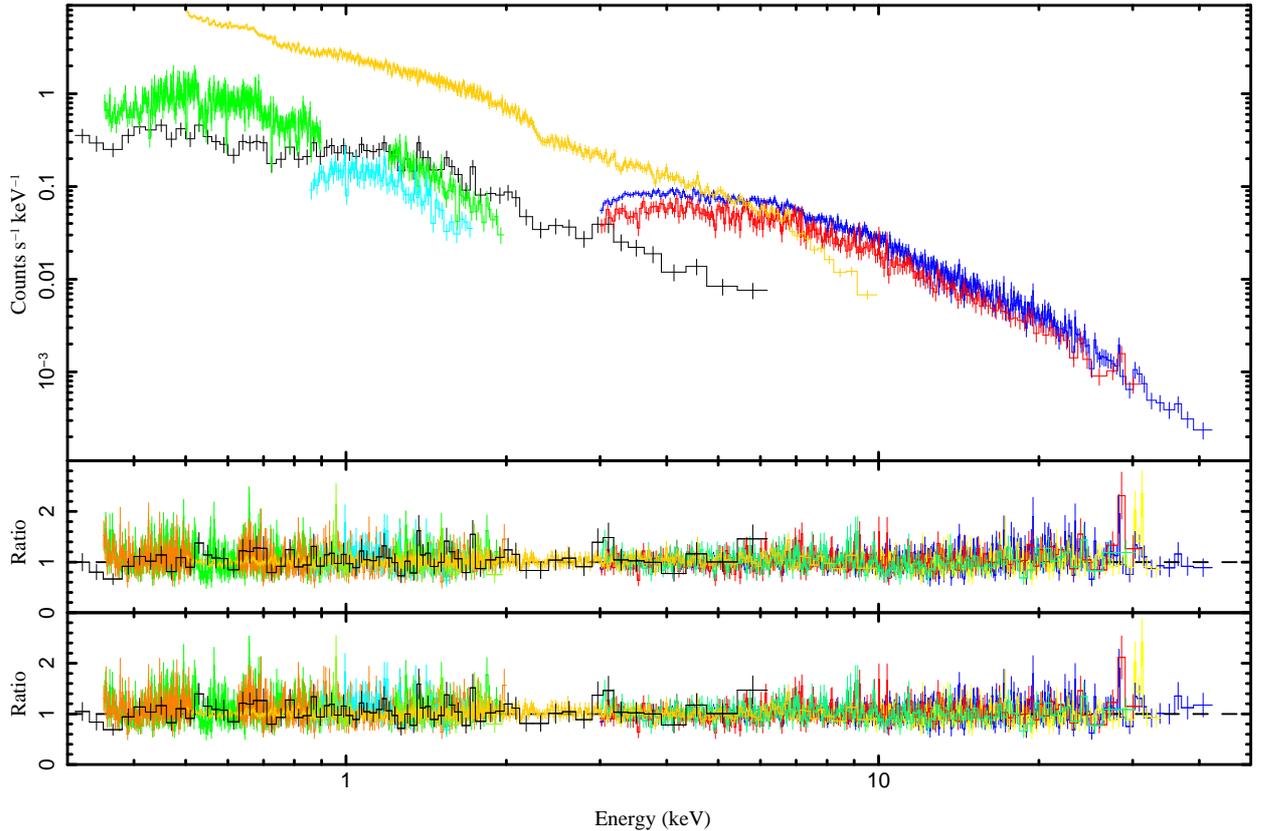}
\caption{Data and residuals of best fitting broadband models. Top: data; middle: residuals for reflection model; bottom: residuals for partial covering model.}
\label{fig:fitres}
\end{figure*}

\subsubsection{Reflection models}
\label{sub:refl}

We first consider the reflection interpretation of the iron line and Compton hump.
Combining the components found in each energy band results in a model of the form 
\texttt{TBnew* (warmabs(1) * warmabs(2) * relxill + photemis)}.

Fits to each observation separately are given in Table~\ref{tab:refl}. The parameters are largely consistent with those found in the individual band models.

The \xmm/\nustar\ observation shows evidence of more emission coming from very close to the black hole than the \swift/\nustar\ observation. This is reflected in the emissivity index and reflection fraction being higher, which can both be induced by light-bending of radiation from a corona close to the black hole \citep{Miniutti04}.

The \swift/\nustar\ observation does not constrain the black hole spin, whereas the \xmm/\nustar\ observation prefers $a>0.87$. The much stronger constraint in the \xmm/\nustar\ observation is largely due to the much greater soft-band (<10\,keV) signal from the \xmm\ coverage. The remaining parameters of this observation also suggest that more emission is from the innermost region, which is most sensitive to the spin. The spin constraint from the \xmm/\nustar\ observation is significantly higher than the fit to the high energy data only ($a<0.43$). This is likely to be because the spin measurement is significantly influenced by the shape of the soft excess and not just the profile of the iron line; the inconsistency could be due to other factors influencing the soft excess \citep[as was found in e.g.][]{parker18spin}, such as the hybrid model discussed later.

The iron abundance of the \xmm/\nustar\ observation is also higher. Since the material in the disc is not expected to change in the 6 months between observations, this is likely to be due to degeneracy with other parameters. If the difference is real, it could be caused by a higher iron abundance in the inner region of the disc, which the higher emissivity index suggests provides more of the reflected component in this observation. However, the data presented here are insufficient to prove this.

The cut-off of the powerlaw is too high to be measured with the current dataset and our lower limits are well outside the observed bandpass. The stronger limit for the \textit{Swift/NuSTAR} observation is due to its greater high-energy signal from the longer \nustar\ exposure.

In order to improve constraints on the parameters of the model which do not change over the 6 month interval between the two observations and increase physical self-consistency, we also perform a joint fit with these parameters -- black hole spin, $a$, accretion disc inclination, $i$, and iron abundance, $A_{\rm Fe}$ -- tied between the two observations. Since the \swift\ spectrum does not significantly constrain the parameters of the absorption and emission, we also tie these parameters between the two observations. This fit is shown in Table~\ref{tab:refl} and Fig~\ref{fig:fitres}. The model spectrum is shown in Fig.~\ref{fig:models}.
The parameters of this fit are largely consistent with the fits to the individual observations. Differences are likely due to parameter degeneracies which are not evident in single-parameter confidence intervals.

\begin{table*}
\caption{Parameters of fits to all data from each observation with reflection model (\texttt{TBnew *  (warmabs(1) * warmabs(2) * relxill + photemis)}). Unconstrained parameters are allowed to vary within the ranges: $-0.998<a<0.998$; $1<\log(\xi/{\rm erg\,cm\,s^{-1}})<2.5$; and $N_{\rm H}$/cm$^{-2}<5\times10^{22}$, indicated in the table by square brackets.}
\label{tab:refl}
\begin{tabular}{lllcccc}
\hline
Component & Model & Parameter & \multicolumn{2}{c}{Separate} & \multicolumn{2}{c}{Joint} \\
&&  & \swift/\nustar & {\it XMM}/\nustar & \swift/\nustar & {\it XMM}/\nustar
\\
  \hline
\multirow{9}{1cm}{Relativistic  reflection} & \multirow{9}{*}{(\texttt{relxill})} & Norm &$1.3\pm0.1\times 10^{-4}$&
$9.9_{-0.3}^{+0.1}\times 10^{-5}$&
$1.24_{-0.04}^{+0.06}\times 10^{-4}$&$9.85_{-0.05}^{+0.04}\times 10^{-5}$ \\
&&Emissivity Index&$2.3_{-0.5}^{+0.7}$&$3.3_{-0.2}^{+0.4}$&
$2.4\pm0.2$&$4.5_{-0.4}^{+1.1}$ \\
&&$a$&Unconstrained&$>0.87$&
\multicolumn{2}{c}{$>0.92$} \\
&&$\theta/^\circ$&$47_{-4}^{+8}$&$36_{-2}^{+1}$&
\multicolumn{2}{c}{$46_{-4}^{+1}$} \\
&&$\Gamma$&$2.28_{-0.06}^{+0.08}$&$2.22_{-0.01}^{+0.02}$&
$2.17_{-0.02}^{+0.01}$&$2.23\pm0.02$ \\
&&$\log(\xi/{\rm erg\,cm\,s^{-1}})$&$1.7_{-0.4}^{+0.3}$&$1.4\pm0.2$&
$1.9_{-0.1}^{+0.1}$&$1.33_{-0.15}^{+0.03}$ \\
&&$A_{\rm Fe}$&$1.0_{-0.2}^{+0.8}$&$2.7_{-0.6}^{+0.7}$&
\multicolumn{2}{c}{$2.9_{-0.4}^{+0.7}$} \\
&&$E_{\rm cut}$/keV&$>290$&$>530$&
$>510$&$>740$ \\
&&$R$&$1.4_{-0.3}^{+0.5}$&$1.8_{-0.3}^{+0.4}$&
$0.9_{-0.1}^{+0.2}$&$2.1\pm0.4$ \\\hline
\multirow{4}{1cm}{Ionised absorption}& \multirow{4}{1cm}{(\texttt{warmabs})} &$N_{\rm H}$/cm$^{-2}$&$<4.7\times 10^{21}$&
$2.0_{-0.4}^{+0.6}\times 10^{21}$&
\multicolumn{2}{c}{$1.8_{-0.4}^{+0.5}\times 10^{21}$} \\
&&$\log(\xi/{\rm erg\,cm\,s^{-1}})$&$[1-2.5]$&$1.96_{-0.04}^{+0.07}$&
\multicolumn{2}{c}{$1.97_{-0.06}^{+0.07}$} \\
&&$N_{\rm H}$/cm$^{-2}$&$<4.9\times 10^{21}$&
$2.0\pm0.3\times 10^{21}$&
\multicolumn{2}{c}{$2.1_{-0.2}^{+0.3}\times 10^{21}$} \\
&&$\log(\xi/{\rm erg\,cm\,s^{-1}})$&$[1-2.5]$&$1.31_{-0.04}^{+0.05}$&
\multicolumn{2}{c}{$1.29\pm0.04$} \\\hline
\multirow{3}{1cm}{Ionised emission}& \multirow{3}{1cm}{(\texttt{photemis})} & Norm &$<0.047$&
$2.6_{-2.1}^{+0.4}\times 10^{-4}$&
\multicolumn{2}{c}{$2.5_{-1.9}^{+0.5}\times 10^{-4}$} \\
&&$N_{\rm H}$/cm$^{-2}$&$[<5\times10^{22}]$&$<5\times 10^{21}$&
\multicolumn{2}{c}{$[<5\times10^{22}]$} \\
&&$\log(\xi/{\rm erg\,cm\,s^{-1}})$&$[1-2.5]$&$1.5\pm0.1$&
\multicolumn{2}{c}{$1.5\pm0.1$} \\\hline
\multirow{8}{1cm}{Cross-calibration constant}&&FPMA/XRT
&$1.15_{-0.08}^{+0.10}$&-&
\multicolumn{2}{c}{$1.14\pm0.03$} \\
&&FPMB/XRT&$1.14_{-0.08}^{+0.09}$&-&
\multicolumn{2}{c}{$1.13\pm0.03$} \\
&&RGS1 order 1/PN&-&$1.00\pm0.02$&
\multicolumn{2}{c}{$1.00\pm0.02$} \\
&&RGS2 order 1/PN&-&$1.01\pm0.02$&
\multicolumn{2}{c}{$1.01\pm0.02$} \\
&&RGS1 order 2/PN&-&$0.99\pm0.04$&
\multicolumn{2}{c}{$0.99\pm0.04$} \\
&&RGS2 order 2/PN&-&$0.97\pm0.04$&
\multicolumn{2}{c}{$0.97\pm0.04$} \\
&&FPMA/PN&-&$1.15_{-0.02}^{+0.03}$&
\multicolumn{2}{c}{$1.15_{-0.03}^{+0.02}$} \\
&&FPMB/PN&-&$1.13_{-0.02}^{+0.03}$&
\multicolumn{2}{c}{$1.13_{-0.03}^{+0.02}$} \\\hline
\multicolumn{3}{c}{$\chi^2$ (bins)}& & & 990 (884) & 2262 (2202) \\
\multicolumn{3}{c}{$\chi^2/$d.o.f.}&$974.7/866=1.13$&$2238/2180=1.03$&\multicolumn{2}{c}{$3252/3056=1.06$}\\
 \hline
\end{tabular}
\end{table*}

\subsubsection{Partial covering models}
\label{sub:pcm}

\begin{figure*}
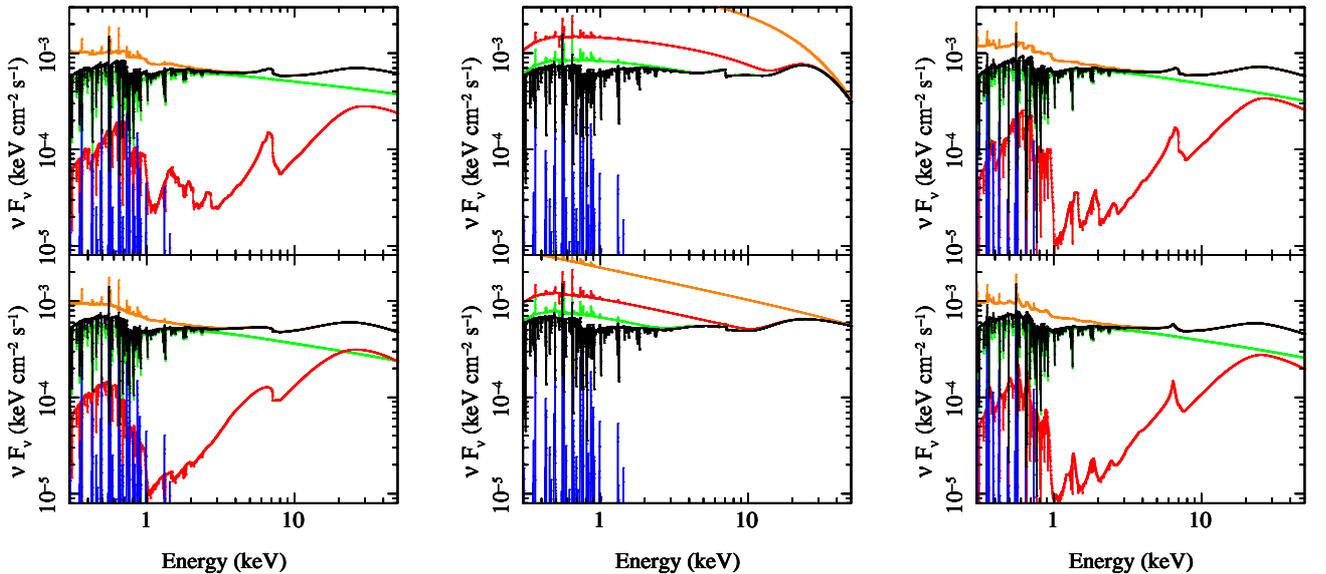

\begin{minipage}{\linewidth}
\includegraphics[width=0.333\linewidth]{modelrefl.ps}
\includegraphics[width=0.333\linewidth]{modelpcf.ps}
\includegraphics[width=0.333\linewidth]{modelhybrid.ps}
\end{minipage}
\caption{Plot of models found for joint fit. Left panels: reflection models; Centre panels: partial covering models; Right panels: hybrid models. Top panels: \swift/\nustar\ observation; Bottom panels: \xmm/\nustar\ observation. Black: total model; Orange: unabsorbed model; Blue: photoionised emission; 
Reflection/hybrid models: Green: powerlaw continuum; Red: reflected component.
Partial covering models: Green: two partial covering components without ionised absorption; Red: one partial covering component.
}
\label{fig:models}
\end{figure*}

We also make a broadband fit with a partial covering model of the form 
\texttt{TBnew * zpcfabs(1) * zpcfabs(2) * (warmabs(1) * warmabs(2) * cutoffpl + photemis)}.
Parameters for this model are given in Table~\ref{tab:pcf}; residuals are shown in Fig.~\ref{fig:fitres} and model spectra in Fig.~\ref{fig:models}.

\begin{table*}
\caption{Parameters of fits to all data from each observation with partial covering model (\texttt{TBnew * zpcfabs(1) * zpcfabs(2) * (warmabs(1) * warmabs(2) * cutoffpl + photemis)}). Unconstrained parameters are allowed to vary within the ranges: $1<\log(\xi/{\rm erg\,cm\,s^{-1}})<2.5$; and $N_{\rm H}$/cm$^{-2}<5\times10^{22}$, indicated in the table by square brackets.}
\label{tab:pcf}\begin{tabular}{lllcccc}
\hline
Component & Model & Parameter & \multicolumn{2}{c}{Separate} & \multicolumn{2}{c}{Joint} \\
&&  & \swift/\nustar & {\it XMM}/\nustar & \swift/\nustar & {\it XMM}/\nustar
\\
\hline
\multirow{4}{1.3cm}{Partial covering absorber}&\multirow{2}{1cm}{(\texttt{zpcfabs})}&$N_{\rm H}$/cm$^{-2}$&$7.2_{-1.1}^{+1.2}\times10^{24}$&$5.5_{-0.6}^{+1.3}\times10^{24}$&
$1.2_{-0.1}^{+0.13}\times10^{25}$&$5.2_{-0.6}^{+1.2}\times10^{24}$ \\
&&$f_{\rm Cov}$&$0.64_{-0.07}^{+0.03}$&$0.51_{-0.03}^{+0.05}$&
$0.69_{-0.06}^{+0.05}$&$0.51_{-0.03}^{+0.05}$ \\
&\multirow{2}{1cm}{(\texttt{zpcfabs})}&$N_{\rm H}$/cm$^{-2}$&$6.2_{-1.1}^{+0.7}\times10^{23}$&$2.0_{-0.2}^{+0}\times10^{23}$&
$1.0\pm0.15\times10^{24}$&$1.9_{-0.2}^{+0}\times10^{23}$ \\
&&$f_{\rm Cov}$&$0.41_{-0.03}^{+0.04}$&$0.361_{-0.003}^{+0}$&
$0.43\pm0.06$&$0.356_{-0.003}^{+0.018}$ \\\hline
\multirow{3}{1cm}{Primary cut-off powerlaw} & \multirow{3}{1cm}{(\texttt{cutoffpl})}& Norm&$0.052_{-0.006}^{+0.014}$&$0.028\pm0.001$&
$0.06\pm0.01$&$0.0276_{-0.0023}^{+0.0001}$ \\
&&$\Gamma$&$2.14_{-0.08}^{+0.10}$&$2.33_{-0.004}^{+0.02}$&
$2.12\pm0.06$&$2.33\pm0.02$ \\
&&$E_{\rm Cut}$/keV&$26_{-6}^{+0}$&$>190$&
$22_{-5}^{+7}$&$>180$ \\\hline
\multirow{4}{1cm}{Ionised absorption}& \multirow{4}{1cm}{(\texttt{warmabs})} &$N_{\rm H}$/cm$^{-2}$&$<8.8\times 10^{21}$&$3.3_{-0.2}^{+0}\times 10^{21}$&
\multicolumn{2}{c}{$3.16_{-0.17}^{+0}\times 10^{21}$} \\
&&$\log(\xi/{\rm erg\,cm\,s^{-1}})$&$>1.2$&$2.06_{-0.05}^{+0.03}$&
\multicolumn{2}{c}{$2.05_{-0.05}^{+0.03}$} \\
&&$N_{\rm H}$/cm$^{-2}$&$<4.6\times 10^{21}$&$2.30_{-0.06}^{+0}\times 10^{21}$
&\multicolumn{2}{c}{$2.29_{-0.06}^{+0}\times 10^{21}$} \\
&&$\log(\xi/{\rm erg\,cm\,s^{-1}})$&$[1-2.5]$&$1.39\pm0.03$&
\multicolumn{2}{c}{$1.40_{-0.05}^{+0.04}$} \\\hline
\multirow{4}{1cm}{Ionised emission}& \multirow{4}{1cm}{(\texttt{photemis})} & Norm&$<1.1\times 10^{-2}$&
$3.4_{-2.4}^{+0.6}\times 10^{-4}$&
\multicolumn{2}{c}{$1.6_{-1.5}^{+0.3}\times 10^{-3}$} \\
&&$N_{\rm H}$/cm$^{-2}$&$[<5\times10^{22}]$&$<1.7\times 10^{21}$&
\multicolumn{2}{c}{$<1.3\times 10^{21}$} \\
&&$\log(\xi/{\rm erg\,cm\,s^{-1}})$&$>1.3$&$1.55_{-0.08}^{+0.07}$&
\multicolumn{2}{c}{$1.53\pm0.08$} \\\hline
\multirow{8}{1cm}{Cross-calibration constant}&&FPMA/XRT&$1.14_{-0.08}^{+0.09}$&-&
\multicolumn{2}{c}{$1.14_{-0.05}^{+0.08}$} \\
&&FPMB/XRT&$1.13_{-0.08}^{+0.09}$&-&
\multicolumn{2}{c}{$1.13_{-0.05}^{+0.08}$} \\
&&RGS1 order 1/PN&-&$1.00\pm0.02$&
\multicolumn{2}{c}{$0.96\pm0.02$} \\
&&RGS2 order 1/PN&-&$1.00\pm0.02$&
\multicolumn{2}{c}{$0.95\pm0.02$} \\
&&RGS1 order 2/PN&-&$0.97\pm0.04$&
\multicolumn{2}{c}{$0.92\pm0.04$} \\
&&RGS2 order 2/PN&-&$0.96\pm0.04$&
\multicolumn{2}{c}{$0.93\pm0.04$} \\
&&FPMA/PN&-&$1.12_{-0.03}^{+0.02}$&
\multicolumn{2}{c}{$1.12_{-0.03}^{+0.02}$} \\
&&FMPB/PN&-&$1.10_{-0.03}^{+0.02}$&\multicolumn{2}{c}{$1.10_{-0.03}^{+0.02}$}
\\
 \hline
\multicolumn{3}{c}{$\chi^2$ (bins)}& & & 980 (884) & 2339 (2202) \\
\multicolumn{3}{c}{$\chi^2/$d.o.f.}&$974.2/868=1.12$&$2297/2182=1.05$&\multicolumn{2}{c}{$3319/3057=1.09$}\\
 \hline
\end{tabular}

\end{table*}

While this produces an acceptable fit to the spectrum, some parameters are not physically likely. In particular, the high-energy cut-off of $22_{-5}^{+7}$\,keV in the \swift/\nustar\ observation is much lower than is typically found in AGN \citep[e.g.][]{fabian15,lubinski16} and below the lowest found so far with \nustar\ data ($42\pm3$\,keV in Ark~564, \citealt{kara17}). The time-averaged \swift-BAT spectrum does not show such a low cut-off energy \citep[e.g.][]{Vasudevan10,ricci17}, although the coronal temperature may change with time. This is corroborated by the low cut-off not being present in the \xmm/\nustar\ observation, which is detected only up to 35\,keV, below the far side of the Compton hump. Forcing the cut-off to be at least 100\,keV results in a significantly worse fit ($\Delta\chi^2=18.5$). 
The low cut-off value may be due to curvature from the high-energy side of the Compton hump being accounted for by an artificially low cut-off energy. A high column density component ($N_{\rm H}=1.2_{-0.1}^{+0.13}\times10^{25}$\,cm$^{-2}$) is then required to produce the low-energy side of the Compton hump.

The large absorbing column density also implies a high unabsorbed luminosity: $L_{\rm 0.5-10\,keV}=7.5\times10^{43}$\,erg\,s$^{-1}$ for the \swift/\nustar\ observation. This is not compatible with the bolometric luminosity of $10^{44}$\,erg\,s$^{-1}$ found by SED fitting \citep{Vasudevan10} which must also include significant intrinsic disc emission.

\begin{figure*}
\begin{minipage}{\linewidth}
\includegraphics[width=0.5\linewidth]{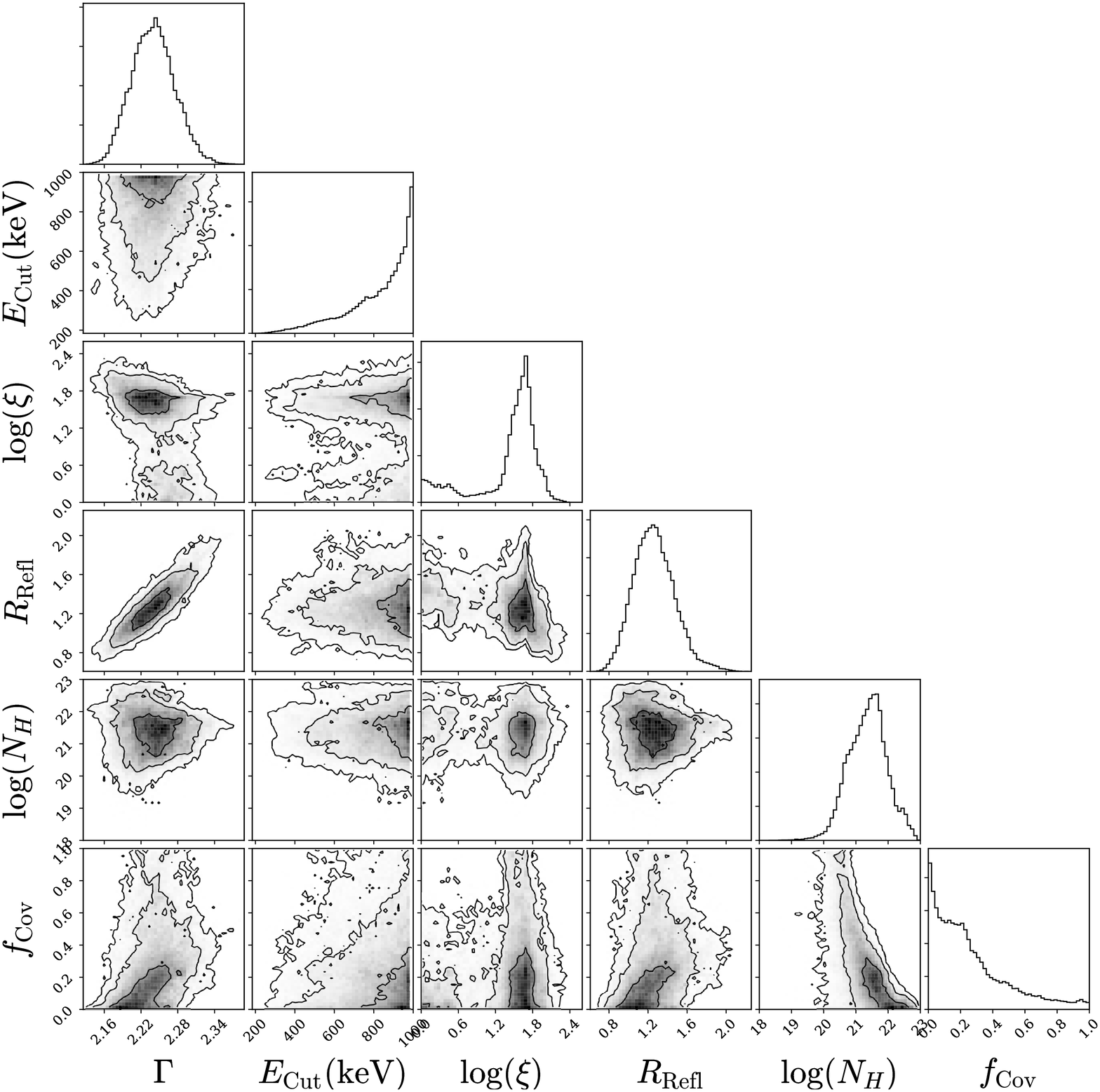}
\includegraphics[width=0.5\linewidth]{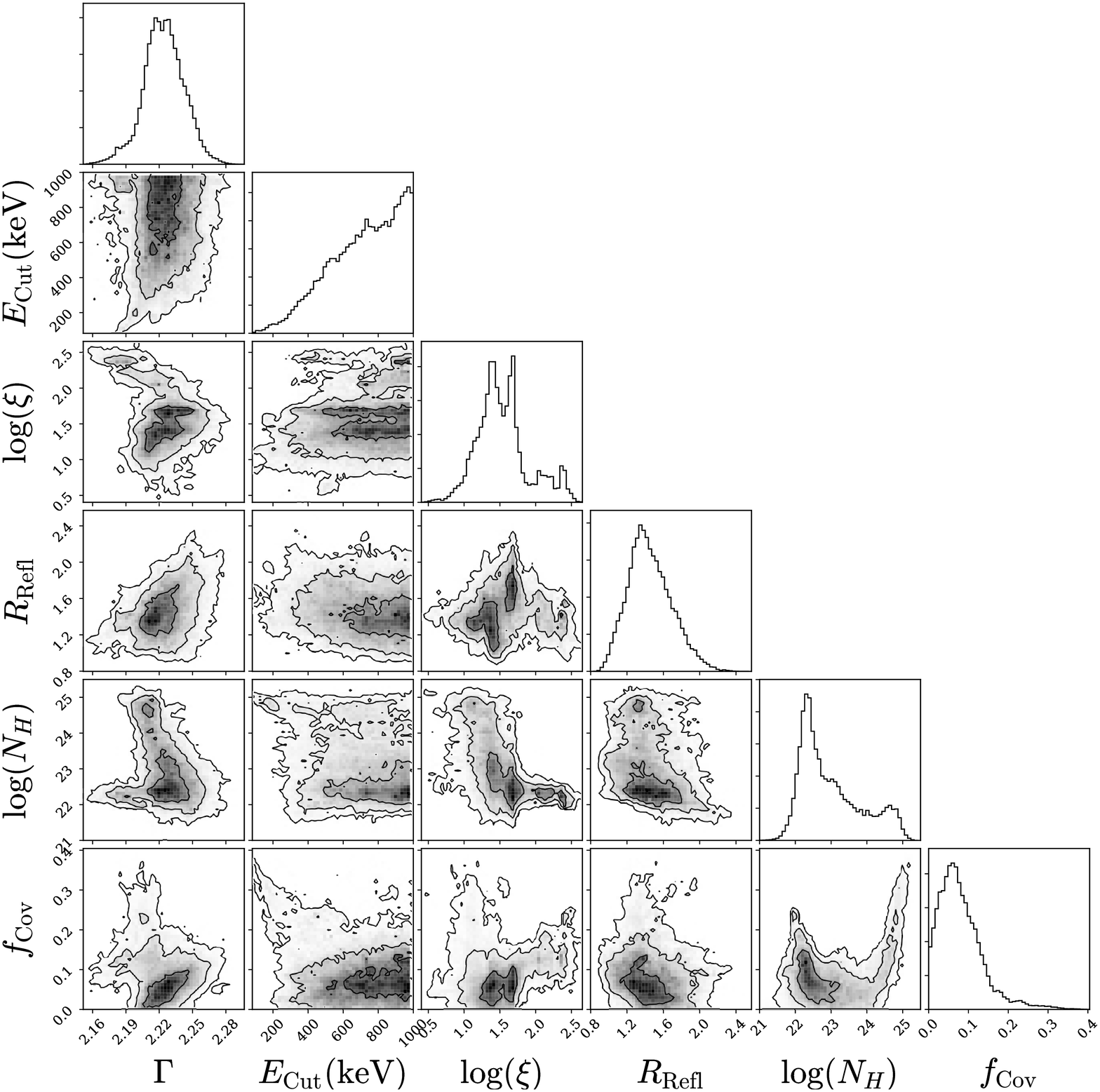}

\includegraphics[width=0.5\linewidth]{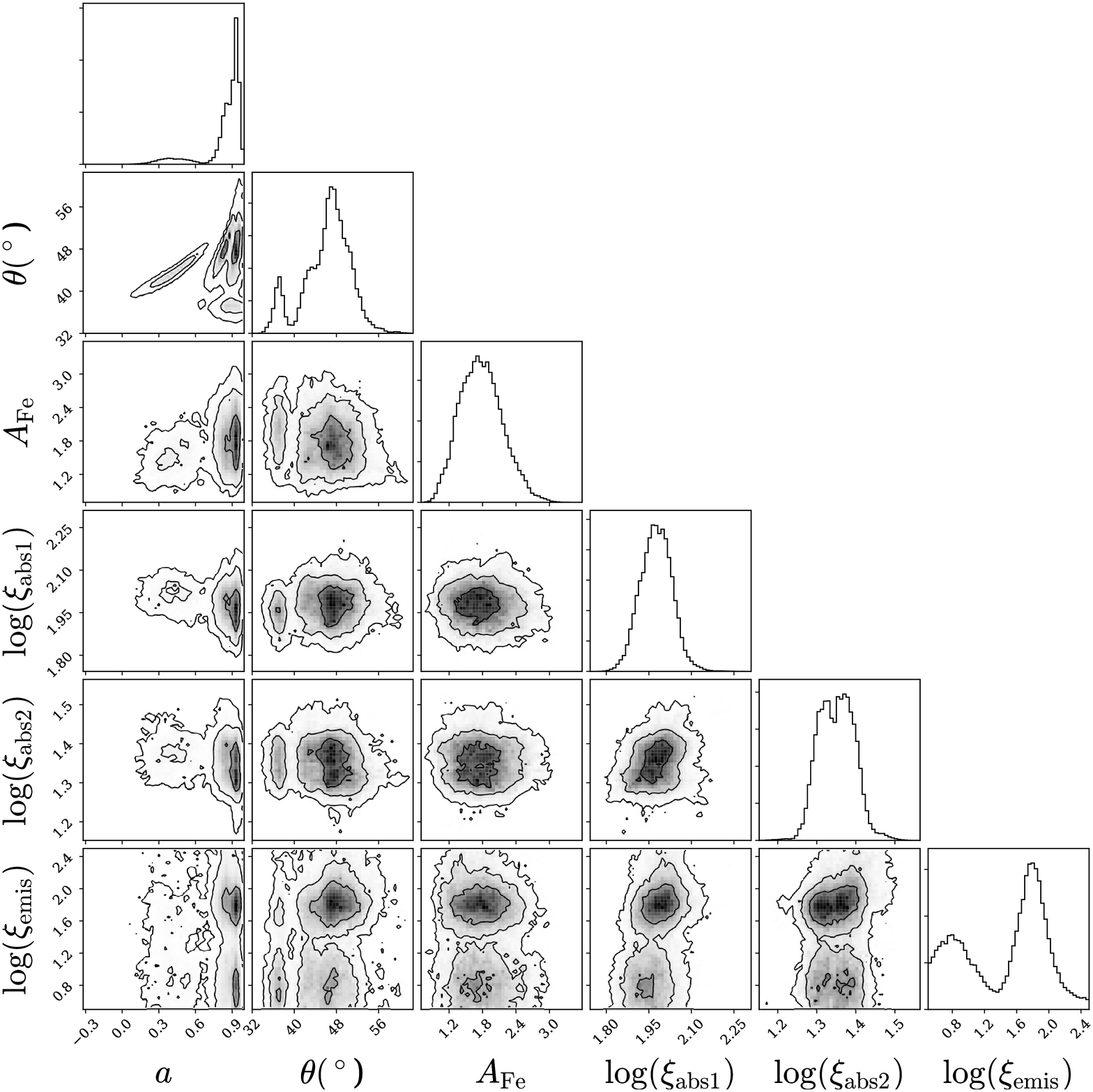}
\includegraphics[width=0.5\linewidth]{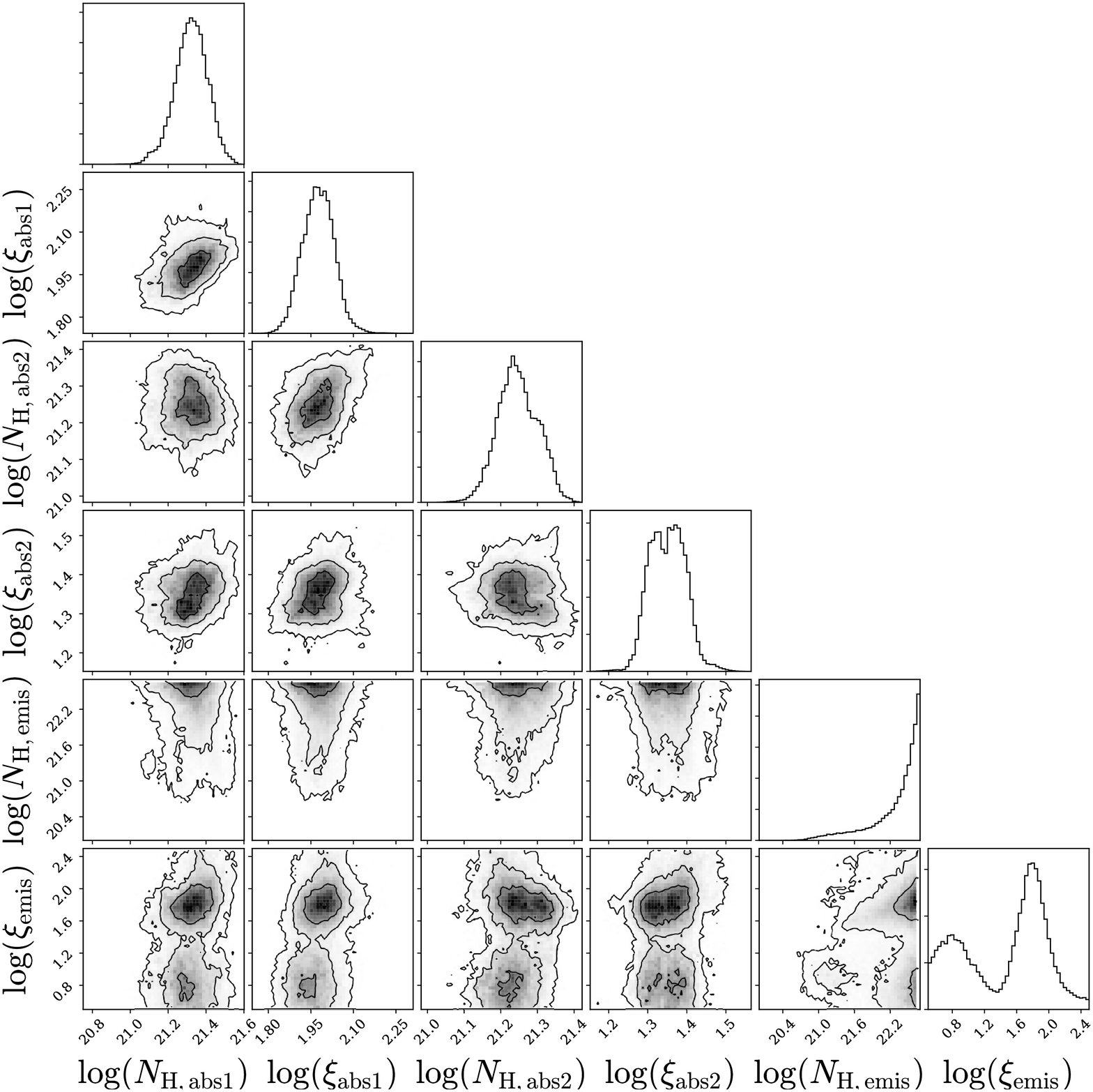}
\end{minipage}
\caption{Corner plots of MCMC parameter estimation for a hybrid model with reflection and partial covering. Parameters shown in each panel are: top left: \swift/\nustar\ observation; top right: \xmm/\nustar\ observation; bottom left: parameters tied between observations; bottom right: warm absorption/emission. Column densities have units of cm$^{-2}$; $\xi$ has units of erg\,cm\,s$^{-1}$. Contours indicate 1,2, and 3$\sigma$ intervals.}

\label{fig:hybrid}
\end{figure*}

\subsubsection{Hybrid models}
\label{sub:hybrid}
Further to the extreme scenarios in which only reflection or only absorption are responsible for the spectrum of Mrk~766, we consider a model which includes both of these effects.
We use a model of the form 
\texttt{TBnew * zpcfabs(1) * (warmabs(1) * warmabs(2) * relxill + photemis)}.
Due to the potential for strong degeneracy between the two possible causes (relativistic emission and partial covering absorption) for the spectral shape, we use Monte-Carlo methods to sample the parameter space. We use the \textsc{XSPEC\_EMCEE} code\footnote{Written by Jeremy Sanders, based on the \textsc{EMCEE} package \citep{foreman13}.}.
We use 600 walkers and take probability densities from 10000 iterations after the chain has converged. Results are shown in table~\ref{tab:hybrid} and Fig.~\ref{fig:hybrid}.

Note that we have shown the $\chi^2$ value from direct minimisation for comparison with other models; the parameters from the two methods are consistent.

The reflection parameters found are largely consistent with those found for the pure reflection case, with less strict confidence limits. The iron abundance is somewhat lower, being slightly closer to Solar, and the reflection fraction of the \xmm/\nustar\ observation is closer to unity. The constraint on the spin is significantly weaker.
The parameters of the absorption are much less extreme than the pure partial covering model. We now find solutions with column densities $N_{\rm H}\simeq 10^{22-23}$\,cm$^{-2}$, which is more reasonable given the lack of a strong narrow iron line, which would be expected with the higher columns required for the pure partial covering model.
In general, the hybrid model requires less extreme parameters than either reflection or partial covering alone.

\begin{table*}
\caption{Parameters of fits to all data from each observation with model including partial covering and reflection 
\texttt{TBnew * zpcfabs(1) * (warmabs(1) * warmabs(2) * relxill + photemis)}.
Parameters given are the median of the posterior distribution; errors correspond to the central 90\% of the MCMC posterior distribution.
The $\chi^2$ value refers to the value found by $\chi^2$-minimisation.
Unconstrained parameters are allowed to vary within the ranges: $1<\log(\xi/{\rm erg\,cm\,s^{-1}})<2.5$; and $N_{\rm H}$/cm$^{-2}<5\times10^{22}$, indicated in the table by square brackets.}
\label{tab:hybrid}
\begin{tabular}{lllcc}
\hline
Component & Model & Parameter &  \multicolumn{2}{c}{Joint} \\
 && &  \swift/\nustar & {\it XMM}/\nustar \\
\hline
\multirow{9}{1cm}{Relativistic reflection}&\multirow{9}{1cm}{(\texttt{relxill})}& 
Norm&$1.30\pm0.04\times 10^{-4}$&$1.02_{-0.03}^{+0.06}\times 10^{-4}$ \\
&& Emissivity Index&$2.2_{-0.6}^{+0.4}$&$5.5_{-1.0}^{+1.7}$ \\
&& $a$ & \multicolumn{2}{c}{$>0.4$} \\
&& $\theta$ & \multicolumn{2}{c}{$47_{-10}^{+6}$} \\
&& $\Gamma$ & $2.24\pm0.06$ & $2.22\pm0.03$ \\
&& $\log(\xi/{\rm erg\,cm\,s^{-1}})$ & $1.6_{-1.4}^{+0.4}$&$1.5_{-0.5}^{+0.8}$ \\
&& $A_{\rm Fe}$&\multicolumn{2}{c}{$1.8_{-0.6}^{+0.7}$} \\
&& $E_{\rm cut}$ &$>490$&$>350$ \\
&& $R$& $1.25_{-0.3}^{+0.4}$&$1.4_{-0.3}^{+0.4}$ \\
\hline
\multirow{2}{2.1cm}{Partial covering absorption}&\multirow{2}{1cm}{(\texttt{zpcfabs})}& 
$N_{\rm H}$/cm$^{-2}$&$<2.3\times10^{22}$&$5_{-1}^{+500}\times10^{22}$\\
&& $f_{\rm Cov}$&$<0.75$&$0.07_{-0.06}^{+0.12}$ \\
\hline
\multirow{4}{1cm}{Ionised absorption}&\multirow{4}{1cm}{(\texttt{warmabs})}& 
$N_{\rm H}$/cm$^{-2}$&\multicolumn{2}{c}{$2.1_{-0.6}^{+0.8}\times 10^{21}$} \\
&& $\log(\xi/{\rm erg\,cm\,s^{-1}})$&\multicolumn{2}{c}{$1.98\pm0.09$} \\
&&$N_{\rm H}$/cm$^{-2}$&\multicolumn{2}{c}{$1.8_{-0.3}^{+0.4}\times 10^{21}$} \\
&& $\log(\xi/{\rm erg\,cm\,s^{-1}})$&  \multicolumn{2}{c}{$1.35\pm0.07$} \\
\hline
\multirow{3}{1cm}{Ionised emission} & \multirow{3}{1cm}{(\texttt{photemis})}&
Norm&\multicolumn{2}{c}{$1.2_{-0.8}^{+0.3}\times 10^{-5}$} \\
&&$N_{\rm H}$/cm$^{-2}$&\multicolumn{2}{c}{$[<5\times10^{22}]$} \\
&& $\log(\xi/{\rm erg\,cm\,s^{-1}})$& \multicolumn{2}{c}{$1.6_{-1.0}^{+0.5}$} \\
\hline
\multicolumn{3}{c}{$\chi^2$ (bins)}& 979 (884) & 2267 (2202) \\
\multicolumn{3}{c}{$\chi^2/$d.o.f.}&\multicolumn{2}{c}{$3246/3052=1.06$}\\
 \hline
\end{tabular}

\end{table*}

\subsection{Variability}
\label{sec_timing}

\begin{figure}
\centering
\includegraphics[width=\linewidth]{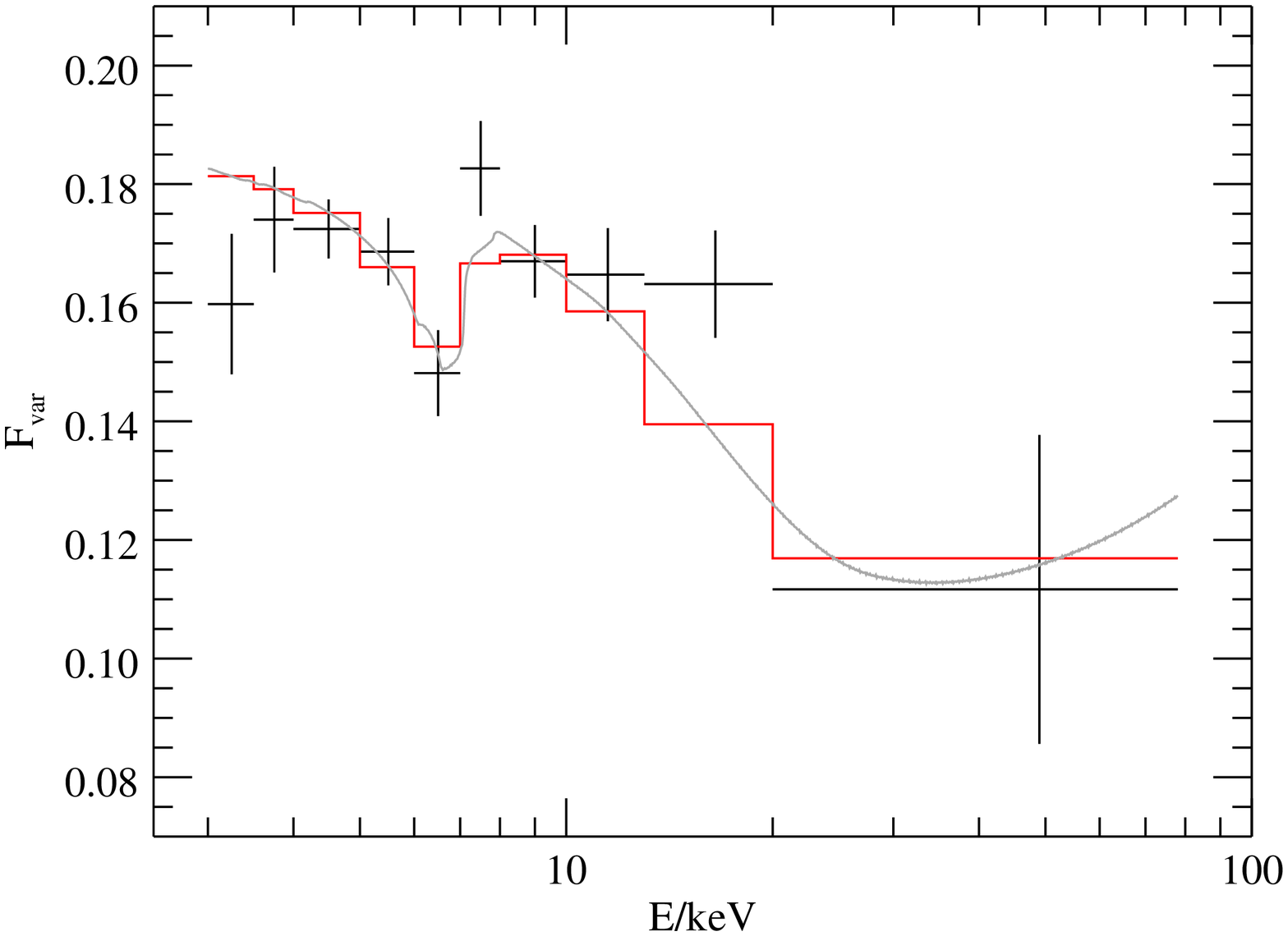}
\caption{
The fractional excess variance for Mrk~766 computed using the entire light curve of the longer observation in time bins of 400\,s.
The variability drops in the iron K and Compton hump regions.
This can be explained by a variable continuum and a constant reflection component (grey; binned to data resolution in red).
}
\label{fig_nustarrms}
\end{figure}

To characterize the rapid variability in this source, we compute the fractional excess variance in the \nustar\ spectrum, binned to 400\,s, 
using the prescription in \citet{edelson02} with errors from the formula in \citet{Vaughan03_variability}. We use the fractional variability rather than Fourier methods owing to the former's insensitivity to the orbital gaps in \nustar\ observations \citep[e.g.][]{nandra97var}. The shorter \xmm/\nustar\ observation only has enough signal to produce two energy bins, which are broadly consistent with the results from the longer observation, so we only present detailed results from the longer observation here (Fig.~\ref{fig_nustarrms}). 

The variability appears reduced in the $6-7$\,keV band, the rest energy of iron~K, and above 10~keV, where the Compton reflection hump provides significant flux in the mean spectrum. 
 Overall, the decrease in variability on short timescales appears to follow the profile of the relativistic reflection features, which suggests that the reflection component does not vary as much as the continuum emission in Mrk~766.  This is reminiscent of previous spectral-timing results of MCG--6-30-15 that also showed that the continuum varies more than the ionised reflection \citep{vaughan03, fabian03, Marinucci14}.
The increase in variability in the $7-8$\,keV bin may be due to variable blueshifted absorption \citep{risaliti11,parker17var}.

We can test whether this $F_{\rm Var}$ spectrum is compatible with a variable continuum and less variable reflection by considering the extreme case in which the reflection is constant, so only the power law varies.
Denoting the mean reflected flux as $f_{\rm R}(E)$, the mean continuum as $f_{\rm C}(E)$ and the variance as $\alpha^2f_{\rm C}^2$ (i.e. the variability is a constant multiple of the continuum flux), the fractional variability is simply $$ F_{\rm Var}(E)=\alpha\frac{f_{\rm C}(E)}{f_{\rm C}(E)+f_{\rm R}(E)} .$$
Thus, higher $\alpha$ leads to higher $F_{\rm Var}$ while the shape of $F_{\rm Var}$ is set by the shape of the reflected spectrum, which we take from the fit to the mean spectrum.
We may then substitute for the mean fluxes found in our fits to the mean spectrum and fit the observed $F_{\rm Var}(E)$ values (by $\chi^2$-minimisation) to find $\alpha$.
This gives an acceptable fit, $\chi^2/{\rm d.o.f.}=16/9$, $p=0.07$.

In the partial covering scenario, the $>20$\,keV bin shows intrinsic continuum variability (since there is little absorption at these energies) while the greater variability at lower energies is due to variable absorption strength or continuum pivoting.
We do not attempt to fit this as covariance between different variability mechanisms rapidly introduces more free parameters than can be constrained by the available data.

\section{Discussion}
\label{section_discussion}

\subsection{The reflection model}

We note that the presence of distant cold reflection is not required in either observation.
This may be due to the relativistic reflection already including a significant component of weakly blurred reflection from a low emissivity index ($2.4\pm0.2$ in the \swift/\nustar\ observation). The lack of a distinct distant component could also be due to absorption reducing the X-ray flux to distant regions of the disc and to the torus, weakening the cold reflection.  Since the warm absorber in our line of sight ($\sim 50^\circ$ from the disc plane) has only a small effect on the transmitted flux, this mechanism would require more absorption in the line of sight of the outer disc and torus. Such absorption would also reduce the correlation between X-ray emission and reprocessing from the disc, consistent with the non-detection of correlated UV/X-ray lags in \citet{buisson17}.
The warm absorption also complicates the spectrum, increasing the uncertainty on the amount of cold reflection and so reducing the significance of any detection.

The lower limit on the cut-off of the primary continuum is higher than is typically found in AGN \citep{fabian15,lubinski16}. However, the limit presented here is far outside the \nustar\ bandpass so is strongly affected by the shape of the reflected part of the spectrum rather than being a direct measurement of the cut-off \citep{garcia15}. Due to the complex absorption in Mrk~766, these features are harder to accurately isolate and so the cut-off may be significantly lower than the statistical limit presented here. The difference in curvature below the cut-off energy between a cut-off powerlaw and a true Comptonisation spectrum may also allow for a lower coronal temperature \citep{furst16}.

\subsection{Comparison with a similar source, MCG--6-30-15}

MCG--6-30-15 is a very well studied AGN with similar spectral appearance to Mrk~766 (and similar Principal Components of variability, \citealt{Parker15}); we therefore compare our interpretations with those found for MCG--6-30-15. 
\citet{Marinucci14} study spectral variability in the available \nustar\ data, considering both reflection and partial covering models. The higher count rate of MCG--6-30-15 allows the observation to be cut into 11 sections.
In the reflection interpretation, $a=0.91_{-0.07}^{+0.06}$, $i=33\pm3^\circ$, similar to values found here and in previous works for Mrk~766 \citep[e.g.][]{rgsrellines}. 
 
A detailed analysis of grating spectra of MCG--6-30-15 from RGS by \citet{sako03,turner04} and from \chandra\ by \citet{holczer10} shows two intrinsic absorption systems with distinct velocities, outflowing at $100\pm50$ and $1900\pm150$\,km\,s$^{-1}$.
The absorption considered in our model is comparable to the lower speed component (in our case with fixed outflow velocity of 350\,km\,s$^{-1}$ based on archival line positions) while the fast component in MCG--6-30-15 is too highly ionised ($\log(\xi/{\rm erg\,cm\,s^{-1}})=3.82\pm0.03$) to have a noticeable effect in the short RGS exposure studied here.

The slow component in MCG--6-30-15 has an ionisation parameter ranging from $\log(\xi/{\rm erg\,cm\,s^{-1}})=-1.5$ to $\log(\xi/{\rm erg\,cm\,s^{-1}})=3.5$, while the two ionisations we consider have $\log(\xi/{\rm erg\,cm\,s^{-1}})\sim1.3$ and $1.9$. \citet{holczer10} also find little gas with $0.5<\log(\xi/{\rm erg\,cm\,s^{-1}})<1.5$ so our component with $\log(\xi/{\rm erg\,cm\,s^{-1}})\sim1.3$ may reflect a mixture of more and less ionised gas.
This is plausible since the CCD spectra of MCG--6-30-15 may be described with a two-state warm absorber with ionisation parameters of $\log(\xi/{\rm erg\,cm\,s^{-1}})\simeq2$ and $\log(\xi/{\rm erg\,cm\,s^{-1}})=1.15-1.65$ \citep{Marinucci14}.

\citet{Emman11} find soft lags in the X-ray light curves of both Mrk~766 and MCG--6-30-15 (as do \citealt{kara14_mcg6}), which are interpreted as arising from the delay of reflected emission.
\citet{Parker14_mcg6} study the variability of MCG--6-30-15 with Principal Component Analysis (PCA), finding that the variability is consistent with that expected from an intrinsically variable X-ray source with less variable relativistic reflection. 
This is corroborated by \citet{miniutti07}, who calculate the RMS (fractional variability) spectrum of a \suzaku\ observation of MCG--6-30-15, which has a similar shape to the short timescale $F_{\rm Var}$ spectrum found here.
This could arise from a vertically extended or two component corona, in which, due to strong light bending, the lower portion principally illuminates the disc while the upper region is responsible for the majority of the direct emission. This would also decorrelate the observed X-ray emission from any UV variability which is driven by disc heating, agreeing with the lack of correlation seen in these two sources \citep{buisson17}.
In the observations presented here, Mrk~766 remains in a high state, so it is hard to find evidence of coronal extension from variability.

\subsection{Distinguishing absorption from reflection}

While variable partial covering absorption and reflection both provide acceptable fits to the data, the reflection model is favoured for the following reasons:

\begin{itemize}

\item the partial covering model gives a high unabsorbed luminosity ($L_{\rm 0.5-10\,keV}=7\times10^{43}$\,erg\,s$^{-1}$), which is incompatible with previous directly integrated measurements of the bolometric luminosity \citep{Vasudevan10}.

\item the high-energy continuum cut-off of $22_{-5}^{+7}$\,keV is very low in the partial covering model of the \swift/\nustar\ observation (although some recent \nustar\ observations have found low cut-offs in other sources, e.g. $53_{-8}^{+11}$\,keV, \citealt{tortosa17}; $42\pm3$\,keV, \citealt{kara17});

\item the PCA analysis in \citet{Parker15} shows that Mrk~766 varies in the same way as MCG--6-30-15, showing the behaviour of a source whose variability is explained well by relativistic reflection from a vertically extended corona;

\item the fractional variability spectrum shows a clear dip in the shape of the iron line, as would be produced by a variable continuum and less variable reflection.

\end{itemize}

\section{Conclusions}
\label{section_conclusions}

We have presented two new observations of Mrk~766 taken by \nustar, providing a detailed view of its hard X-ray spectrum. With simultaneous coverage in soft X-rays by \xmm\ or \swift, we are able to exploit the high spectral resolution of \xmm-RGS to take account of warm absorption and so produce better constraints on the broadband spectrum.

We can model the spectrum with reflection or partial covering to generate the iron~K feature and Compton hump.
In the reflection model, the system has a high spin black hole ($a>0.92$) viewed at intermediate inclination ($i=46_{-2}^{+1}\,^\circ$).
The best-fitting partial covering model is questionable as it requires a very low cut-off energy and the intrinsic X-ray luminosity is high compared to the bolometric luminosity.
A hybrid model including reflection and partial covering allows less extreme conditions for each component of the model.

\section*{Acknowledgements}

DJKB, MLP and RSV acknowledge financial support from the Science and Technology Facilities Council (STFC).
ACF, AML, CP and MLP acknowledge support from the ERC Advanced Grant FEEDBACK 340442.
This work made use of data from the \nustar\ mission, a project led by the California Institute of Technology, managed by the Jet Propulsion Laboratory, and funded by the National Aeronautics and Space Administration. This research has made use of the \nustar\ Data Analysis Software (NuSTARDAS) jointly developed by the ASI Science Data Center (ASDC, Italy) and the California Institute of Technology (USA). This work made use of data supplied by the UK Swift Science Data Centre at the University of Leicester.
This work has made use of observations obtained with \xmm, an ESA science mission with instruments and contributions directly funded by ESA Member States and NASA.
We acknowledge support from the Faculty of the European Space Astronomy Centre (ESAC).

\bibliographystyle{mnras}
\bibliography{paper}

\bsp	% typesetting comment
\label{lastpage}
\end{document}